\documentclass{PoS}

\title{Lattice QCD simulation\\
at finite chiral chemical potential}

\ShortTitle{Lattice QCD simulation at finite chiral chemical potential}

\author{Arata~Yamamoto\\
Department of Physics, The University of Tokyo, Tokyo 113-0033, Japan\\
E-mail: \email{a-yamamoto@nt.phys.s.u-tokyo.ac.jp}}

\abstract{
Chiral chemical potential does not cause the sign problem in the Monte Carlo simulation of lattice QCD.
Using the chiral chemical potential, we study the chiral magnetic effect in two-flavor full QCD.
We show that a strong external magnetic field induces an electric current in a chirally imbalanced QCD matter.
The qualitative feature of the induced current is consistent with an analytical prediction.
}

\FullConference{The XXIX International Symposium on Lattice Field Theory\\
July 10-16, 2011\\
Squaw Valley, Lake Tahoe, California}

\begin{document}

\section{Introduction}
Lattice QCD simulation is a powerful tool for nonperturbative analysis of QCD.
However, it breaks down at a finite baryon chemical potential because of the famous sign problem.
The fermion determinant becomes complex and the naive Monte Carlo sampling is invalid.
For a small chemical potential, many methods has been proposed.
It is difficult to apply such methods to a large chemical potential.
In some special cases, we can exactly avoid the sign problem.
The famous examples are isospin chemical potential and two-color QCD.
These cases are well studied both in phenomenological studies and in lattice simulations.
We here consider another possibility, that is, {\it chiral chemical potential} \cite{Yamamoto:2011gk}.
Compared with the isospin chemical potential and two-color QCD, the chiral chemical potential has been less studied so far.

The chiral chemical potential $\mu_5$ is defined in the Dirac operator as
\begin{eqnarray}
D(\mu_5) = \gamma_\mu D_\mu + m + \mu_5 \gamma_4 \gamma_5
\label{eqaction}
\end{eqnarray}
\cite{Fukushima:2008xe}.
The chiral chemical potential generates a finite chiral charge.
The chiral charge is an imbalance between the left-handed and right-handed fermion numbers.
The important property of the chiral chemical potential is that it does not cause the sign problem.

Using the chiral chemical potential, we analyze the chiral magnetic effect.
The chiral magnetic effect is an electric current induced by a strong magnetic field in a heavy-ion collision \cite{Kharzeev:2008ey}.
A noncentral collision of two heavy ions produces a very strong magnetic field perpendicular to the reaction plane.
The strong magnetic field fixes the spin and momentum directions of the quarks depending on their chiralities. 
The positive-helicity particles and the negative-helicity particles generates electric currents in the opposite direction.
If the left-handed and right-handed chiralities are symmetric, these two contributions exactly cancel out, and the net current is zero.
If the chiralities are imbalanced, the net current is finite.
In QCD, such an imbalance is locally generated by the axial anomaly and the topological fluctuation of the background gauge field.
The chiral magnetic effect is a direct experimental evidence of the topological fluctuation or the event-by-event CP violation \cite{Kharzeev:2004ey,Kharzeev:2007tn}.

In this study, we use the chiral chemical potential, instead of the topological fluctuation.
This is a different approach from other lattice simulations for the chiral magnetic effect \cite{Buividovich:2009wi,Abramczyk:2009gb,Braguta:2010ej}.
The chiral chemical potential generates the chirally imbalanced QCD matter as an equilibrium state.
Strictly speaking, the chiral charge is not conserved quantity, and this prescription is a kind of approximation of real QCD.
However, the chiral chemical potential is quite convenient for theoretical studies.
The chiral chemical potential was introduced in a phenomenological work, and the induced current of the chiral magnetic effect was derived as
\begin{eqnarray}
j = \frac{1}{2\pi^2} \mu_5 qB
\label{eqj3}
\end{eqnarray}
using the Dirac equation coupled with the background magnetic field \cite{Fukushima:2008xe}.

\section{Simulation setup}
We performed the two-flavor Hybrid Monte Carlo simulation with the chiral chemical potential.
We used the Wilson gauge action and the Wilson fermion action.
The color number is $N_c=3$.
For sake of simplicity, we considered two fermion flavors with the same mass $m$ and charge $q$.
This approximation simplifies the numerical simulation, especially the hybrid Monte Carlo algorithm.

The Wilson-Dirac operator with the chiral chemical potential is 
\begin{eqnarray}
[D_W(\mu_5)]_{x,y}
&=& \delta_{x,y}
- \kappa \sum_i \Bigl[ (1-\gamma_i)U_i(x) \delta_{x+\hat{i},y}
+ (1+\gamma_i)U^\dagger_i(x-\hat{i}) \delta_{x-\hat{i},y} \Bigr] \nonumber\\
&&- \kappa \Bigl[ (1-\gamma_4 e^{a\mu_5\gamma_5})U_4(x) \delta_{x+\hat{4},y}
+ (1+\gamma_4 e^{-a\mu_5\gamma_5})U^\dagger_4(x-\hat{4}) \delta_{x-\hat{4},y} \Bigr]
\label{eqDirac} \\
e^{\pm a\mu_5\gamma_5} &=& \cosh (a\mu_5) \pm \gamma_5\sinh (a\mu_5) \ .
\end{eqnarray}
This is the simplest choice of the chiral chemical potential in lattice QCD.
The form of the chiral chemical potential is analogous to baryon chemical potential \cite{Hasenfratz:1983ba}.
The Dirac operator satisfies the relation
\begin{eqnarray}
\gamma_5 D(\mu_5) \gamma_5 =  D^\dagger (\mu_5) \ .
\end{eqnarray}
From this relation, we can immediately show that the fermion determinant $\det D(\mu_5)$ is always positive real and thus the sign problem does not occur in the case of even flavors.

For the analysis of the chiral magnetic effect, we introduced an external magnetic field.
On the lattice, the U(1) gauge field is introduced as the Abelian phase factor $u_\mu(x)$.
For the external magnetic field, the SU($N_c$) link variable is replaced as
\begin{eqnarray}
U_\mu(x) \to u_\mu(x) U_\mu(x)
\end{eqnarray}
only in the Dirac operator (\ref{eqDirac}).
The kinetic term of the U(1) gauge field is not introduced in the Lagrangian.
The strength of the magnetic field is an external parameter and is not affected by dynamical effects.
For applying a constant magnetic field $B$ in the $x_3$-direction, the phase factor is set as
\begin{eqnarray}
 u_1(x)&=&\exp(-iaqBN_sx_2) \quad {\rm for}\ x_1=aN_s  \nonumber \\
 u_2(x)&=&\exp(iaqBx_1)\\
 u_\mu(x)&=&1 \quad {\rm for}\ {\rm other}\ {\rm components}\ \nonumber
\end{eqnarray}
with $a^2qB=(2\pi/N_s^2)\times$(integer) \cite{Buividovich:2009wi}.

\section{Chiral magnetic effect}
In this section, the lattice gauge coupling is $\beta=2N_c/g^2=5.32144$ and the hopping parameter is $\kappa=0.1665$.
These values correspond to the lattice spacing $a \simeq 0.13$ fm and the pion mass $m_\pi \simeq 0.4$ GeV \cite{Orth:2005kq}.

\begin{figure}[t]
\begin{center}
\includegraphics[scale=1.2]{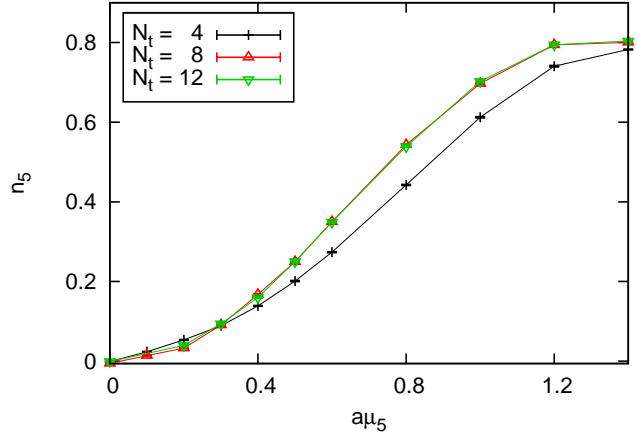}
\caption{\label{fig1}
The chiral charge density $n_5$.
The lattice sizes are $N_s^3 \times N_t = 12^3\times 4, 12^3\times 8$, and $12^4$.
}
\end{center}
\end{figure}

In Fig.~\ref{fig1}, we plot the chiral charge density
\begin{eqnarray}
n_5 \equiv - a^3\langle \bar{\psi} \gamma_4 \gamma_5 \psi \rangle
= a^3\langle \psi^\dagger_L \psi_L - \psi^\dagger_R \psi_R \rangle
\end{eqnarray}
scaled by the lattice unit.
At a finite chiral chemical potential, the chiral charge density is finite, i.e., the system is chirally imbalanced.
The chiral charge density increases as the chiral chemical potential increases.
The qualitative behaviors are different between the confinement phase ($N_t=8$ and 12) and the deconfinement phase ($N_t=4$).
Note, however, that the chiral charge density saturates in $a\mu_5> 1.0$.
This artificial behavior is called as saturation, which is also known in the cases of isospin chemical potential and two-color QCD \cite{Kogut:2002tm}.
The lattice calculation works only below the saturation, i.e.,  $a\mu_5\le  1.0$ in the present setup.

For analyzing the chiral magnetic effect, we applied the external magnetic field to this chirally imbalanced QCD matter.
We calculated the vector current density
\begin{eqnarray}
j_\mu \equiv a^3\langle \bar{\psi} \gamma_\mu \psi \rangle \ .
\label{eqlvc}
\end{eqnarray}
We applied the magnetic field in the $x_3$-direction and measured the transverse component $j_1$ and the longitudinal component $j_3$.
The two transverse components are the same, $j_1=j_2$, because of the rotational symmetry.
The simulation was done in the deconfinement phase ($N_t=4$), where the chiral magnetic effect is expected in heavy-ion collisions.

\begin{figure}[t]
\begin{center}
\includegraphics[scale=1]{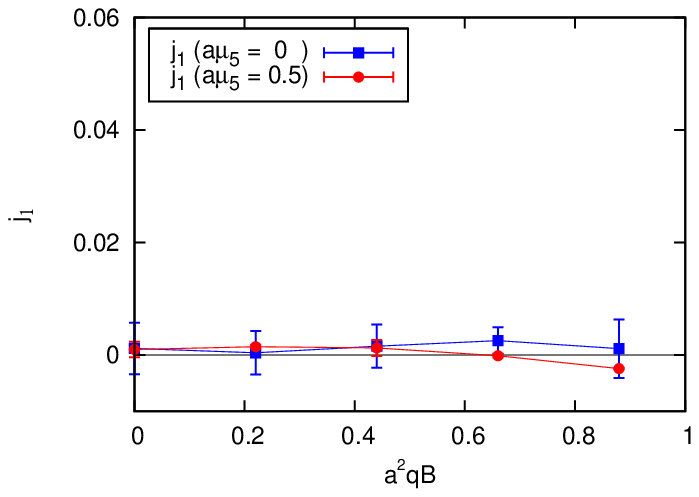}
\includegraphics[scale=1]{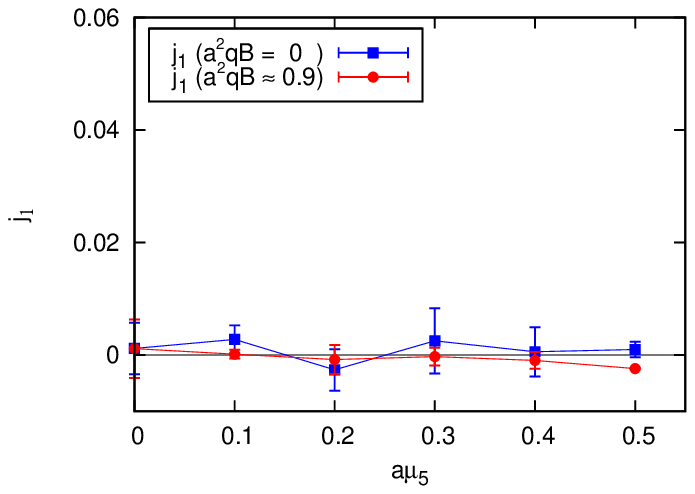}
\caption{\label{fig2}
The transverse component $j_1$ of the vector current density.
The left and right panels are plotted as a function of $qB$ and  $\mu_5$, respectively.
The lattice size is $N_s^3 \times N_t = 12^3\times 4$.
}
\end{center}
\begin{center}
\includegraphics[scale=1]{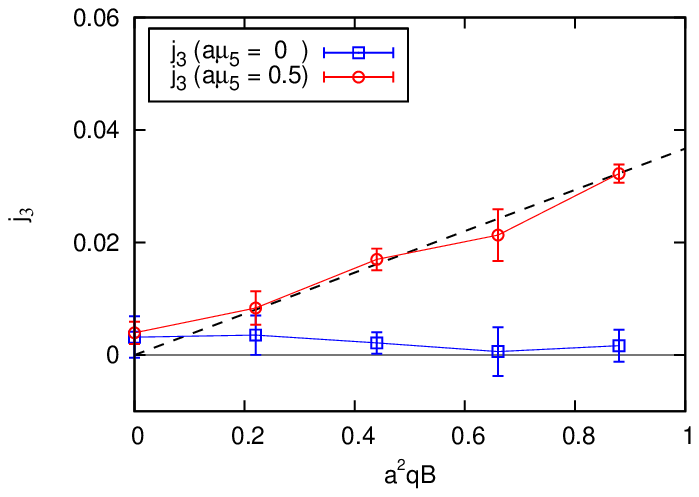}
\includegraphics[scale=1]{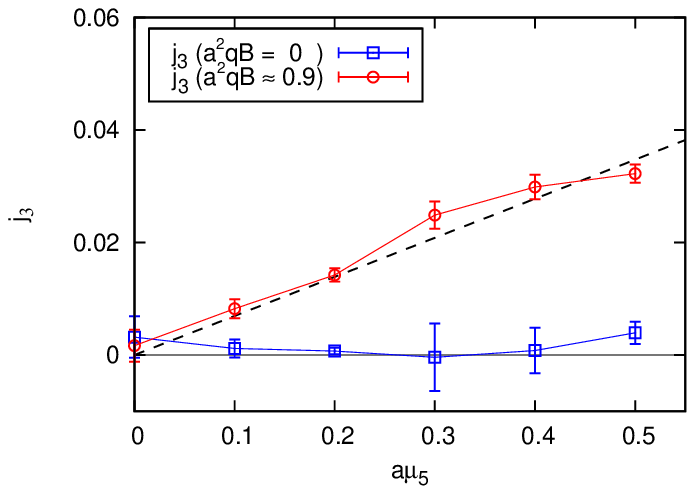}
\caption{\label{fig3}
The longitudinal component $j_3$ of the vector current density.
The notation is the same as in Fig.~2.
The black dashed line is a linear function (3.3).
}
\end{center}
\end{figure}

In Fig.~\ref{fig2}, the transverse component $j_1$ is plotted as a function of the magnetic field $B$ (left) and of the chiral chemical potential $\mu_5$ (right).
The transverse component is always zero because it is irrelevant for the chiral magnetic effect.
As shown in Fig.~\ref{fig3}, the longitudinal component $j_3$ is a linearly increasing function of $B$ at finite $\mu_5$ and a linearly increasing function of $\mu_5$ at finite $B$.
We can parametrize the induced current as
\begin{eqnarray}
j_3 = a^3 C N_{\rm dof} \mu_5 qB \ .
\end{eqnarray}
The factor $N_{\rm dof} = N_c \times N_f =6$ is the number of quarks with the same charge.
This functional form is consistent with the analytical approach (\ref{eqj3}).
The overall constant $C$ is $0.013 \pm 0.001$ in the present lattice simulation.
To compare this value with the analytical approach, we need to calculate the renormalization constant because the local vector current (\ref{eqlvc}) is not renormalization-group invariant.
When we carefully estimate the renormalization effect and systematic errors, we can evaluate QCD corrections to the analytical formula (\ref{eqj3}).
A possible correction has been suggested in a phenomenological work \cite{Fukushima:2010zza}.

In this study, we have succeeded in observing a finite induced current.
This is completely different from other lattice simulations.
In the standard lattice QCD without the chiral chemical potential, the induced current of the chiral magnetic effect cannot be observed \cite{Buividovich:2009wi}.
Since lattice QCD can reproduce nontrivial topological sectors, one might think that it can reproduce the induced current.
However, this is not so easy.
In the usual lattice simulation, the number of the topological charge is typically $O(1)$, and the lattice volume is typically $a^4V \sim O(10^5)$.
The topological charge per volume is thus $O(10^{-5})$.
This is rather small.
In our simulation at a finite chiral chemical potential, the chiral charge density is independent of the volume and its value is $O(10^{-1})$.
Owing to such a large value, we can directly observe the induced current of the chiral magnetic effect.
This is a great advantage of the chiral chemical potential.

\section{Confinement/deconfinement phase transition}

\begin{figure}[t]
\begin{center}
\includegraphics[scale=1.2]{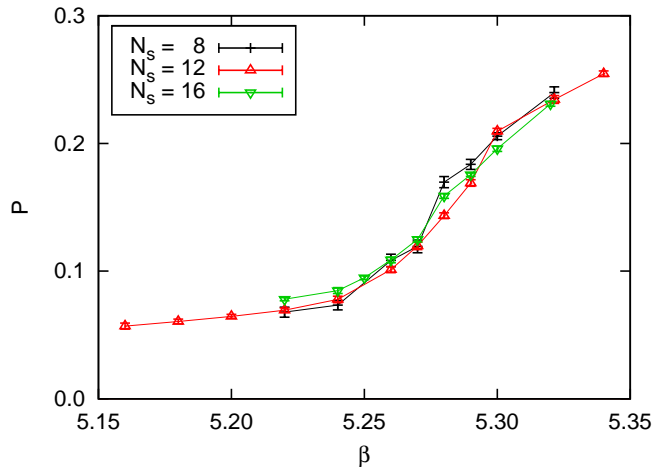}
\caption{\label{fig4}
The expectation value of the Polyakov loop $P$ at $a\mu_5 = 1.0$.
The lattice sizes are $N_s^3 \times N_t = 8^3\times 4, 12^3\times 4$, and $16^3\times 4$.
}
\end{center}
\end{figure}

We also analyzed the phase structure of a chirally imbalanced QCD matter.
The phase structure in $\mu_5$-$T$ plane was studied in phenomenological models \cite{Fukushima:2010fe,Chernodub:2011fr,Ruggieri:2011xc}.
We examined the temperature dependence of the Polyakov loop by varying the lattice gauge coupling $\beta$.
The Polyakov loop in full QCD is a phenomenological criterion for confinement/deconfinement.
The hopping parameter is fixed at $\kappa=0.1665$.
The magnetic field is not applied here.

In Fig.~\ref{fig4}, we plot the Polyakov loop at $a\mu_5 = 1.0$.
The Polyakov loop rapidly rises at $\beta \simeq 5.27$.
This corresponds to a confinement/deconfinement phase transition.
The low-temperature side is the confinement phase, and the high-temperature side is the deconfinement phase.
We determined the order of the phase transition from the dependence on the spatial volume $V=a^3N_s^3$.
As shown in the figure, the Polyakov loop is almost independent of the spatial volume.
We checked that its susceptibility is also independent of the volume.
This scaling behavior suggests that this transition is a crossover.
We also calculated at $a\mu_5 = 0$ and 0.5, and found that the situation is the same.
Thus, we conclude that the order of the confinement/deconfinement phase transition does not change in $0 \le a\mu_5 \le 1.0 $ in the present setup.
Since the nature of the phase transition depends on the quark mass, the calculation at a different quark mass can leads a qualitatively different result, such as a first-order phase transition or a critical endpoint.

\section*{Acknowledgments}
The author thanks K.~Fukushima, T.~Hatsuda, and S.~Sasaki for useful discussions.
This work was supported in part by the Grant-in-Aid for Scientific Research in Japan under Grant No.~22340052.
The lattice QCD simulations were carried out on NEC SX-8R in Osaka University.

\end{document}